\documentclass[conference]{IEEEtran}
\IEEEoverridecommandlockouts
\usepackage{cite}
\usepackage{amsmath,amssymb,amsfonts}

\usepackage{algorithmic}
\usepackage{graphicx}
\usepackage{textcomp}
\usepackage{xcolor}
\def\BibTeX{{\rm B\kern-.05em{\sc i\kern-.025em b}\kern-.08em
    T\kern-.1667em\lower.7ex\hbox{E}\kern-.125emX}}
\begin{document}

\title{\name: Channel-Aware Beam Shaping for Reliable Key Generation in mmWave Wireless Networks
\thanks{This work was supported by the Center for Ubiquitous Connectivity (CUbiC), sponsored by Semiconductor Research Corporation (SRC) and the Defense Advanced Research Projects Agency (DARPA) under the JUMP 2.0 program, award number HR0011-23-3-0002.}}

\author{\IEEEauthorblockN{Poorya Mollahosseini}
\IEEEauthorblockA{Princeton University\\
Princeton, New Jersey 08544\\
Email: poorya@princeton.edu}
\and
\IEEEauthorblockN{Yasaman Ghasempour}
\IEEEauthorblockA{Princeton University\\
Princeton, New Jersey 08544\\
Email: ghasempour@princeton.edu}}

\newcommand{\pmh}[1]{\textcolor{blue}{#1}}

\newcommand{\name} {\mbox{mmKey}}
\newcommand{\yg}[1]{\textcolor{red}{\textbf{#1}}}

\maketitle

\begin{abstract}
Physical-layer key generation (PLKG) has emerged as a promising technique to secure next-generation wireless networks by exploiting the inherent properties of the wireless channel. However, PLKG faces fundamental challenges in the millimeter wave (mmWave) regime due to channel sparsity, higher phase noise, and higher path loss, which undermine both the randomness and reciprocity required for secure key generation. In this paper, we present \name, a novel PLKG framework that capitalizes on the availability of multiple antennas at mmWave wireless nodes to inject randomness into an otherwise quasi-static wireless channel. Different from prior works that sacrifice either the secrecy of the key generation or the robustness, \name~balances these two requirements. In particular, \name~leverages a genetic algorithm to gradually evolve the initial weight vector population toward configurations that suppress the LOS component while taking into account the channel conditions, specifically, the sparsity and the signal-to-noise ratio (SNR). Extensive simulations show that \name~improves the secrecy gap by an average of 39.4\% over random beamforming and 34.0\% over null beamforming, outperforming conventional schemes.

\end{abstract}


\section{Introduction}
\vspace{-1mm}
In recent years, there has been growing interest in adopting cross-layer approaches to secure wireless networks. One promising strategy is to complement traditional symmetric encryption schemes with physical-layer key generation (PLKG)~\cite{zeng2015physical}. PLKG leverages two fundamental properties of the wireless channel that are critical for secret key generation: \emph{randomness} and \emph{reciprocity}. Randomness relies on a rich multipath environment to ensure unpredictability, while reciprocity requires accurate channel measurements at the two legitimate parties, Alice and Bob, to guarantee that both observe highly correlated channels and, hence, keys.

Unfortunately, PLKG faces fundamentally new challenges in the millimeter wave (mmWave) regime compared to the sub-6~GHz band, which undermine both the randomness and reciprocity principles. mmWave channels are inherently sparse~\cite{molisch2016millimeter} due to increased reflection loss and reduced diffraction~\cite{anderson2004building}, making the channel largely a deterministic function of node positions rather than the environment. This line-of-sight (LOS) dominance has two key consequences: (i) for stationary Alice and Bob, the channel becomes static, leading to repetitive and non-random key sequences, and (ii) an adversary, Eve, can accurately estimate the channel simply by observing the LOS path.

Furthermore, while the uplink and downlink channels are theoretically reciprocal, impairments such as noise---whether additive, like additive white Gaussian noise (AWGN), or multiplicative, like phase noise---can degrade reciprocity in practice. To mitigate these effects, a high signal-to-noise ratio (SNR) is required, which is challenging to achieve due to the increased path loss at mmWave frequencies.

\begin{figure}[t]
    \centering
    \includegraphics[width=0.48\textwidth]{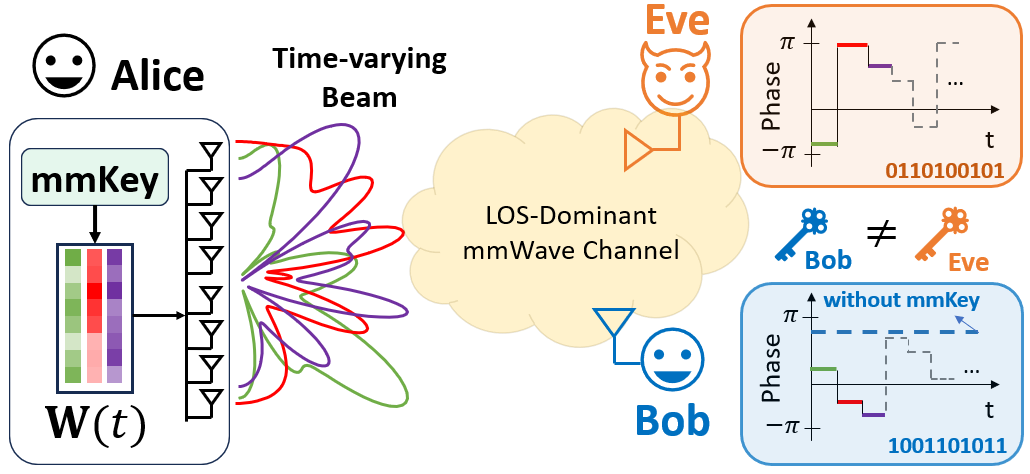}
    \vspace{-3mm}
    \caption{\textbf{System Overview.} \name~generates a series of carefully-designed random-like beam patterns for robust and secure PLKG in sparse mmWave channels.} 
    \label{fig:overview}
     \vspace{-5mm}
\end{figure}

To address the challenges of secure PLKG in the high-frequency mmWave regime, we propose \emph{\name}, a novel channel-aware beam-shaping framework that leverages the multi-antenna capabilities of mmWave hardware to intelligently inject randomness into the channel while enabling robust key extraction under high path loss and phase noise. An overview of the system is shown in Fig.~\ref{fig:overview}. At a high level, dynamically varying the beamforming vector at Alice introduces fluctuations in Bob’s observed channel, which translate into variations in the generated bit sequence.

The goal of \name\ is to induce fluctuations in the observed channel at the legitimate receiver by distributing power and phase across different sparse paths in the environment. This is accomplished by designing a sequence of time-varying weight vectors at the transmitter that jointly satisfy two objectives: \textit{(i) Robustness}—ensuring highly correlated keys between Alice and Bob with minimal mismatches, and \textit{(ii) Secrecy}—ensuring that the eavesdropper’s key is statistically independent. We consider a strong passive adversary with perfect knowledge of Alice and Bob’s locations (and hence their LOS channel), and extremely high receive sensitivity, enabling detection of arbitrarily weak signals without being exposed.

Due to the unique channel characteristics at mmWave frequencies, there exists an inherent tradeoff between robustness and secrecy. The LOS path carries most of the signal power, so steering a highly directive beam toward it enhances robustness by improving key agreement between Alice and Bob. However, this also enables Eve to accurately estimate the effective channel by monitoring LOS variations, compromising secrecy. Alternatively, allocating minimal power to the LOS—e.g., through nulling—improves secrecy by reducing Eve’s channel correlation but severely degrades Bob’s SNR, increasing the key mismatch rate. Furthermore, precise nulling may not be feasible in practice due to the limited phase and amplitude resolution at the transmitter.

In contrast, \name\ balances secrecy and robustness by leveraging prior channel knowledge and accounting for limited phase and amplitude resolution at the transmitter. It employs a genetic algorithm (GA) to efficiently explore the large, discrete beam space and maximize the secrecy gap, defined as the difference between Alice–Bob and Bob–Eve key disagreement rates. The GA evolves pseudo-random beams toward LOS-suppressing solutions while maintaining population diversity. A subset of weight vectors is then selected based on the observed channel conditions (e.g., sparsity and SNR), ensuring that the final beam sequence achieves a desirable tradeoff between secrecy and robustness.

We perform extensive simulations to evaluate the performance of \name~against two baseline methods—--random beamforming and null beamforming—--across a wide range of channels. Overall, our results indicate that \name~offers a flexible and effective solution for enabling secure PLKG in static mmWave environments, achieving an average secrecy gap improvement of 39.4\% and 34.0\% compared to random beamforming and null beamforming, respectively.

\vspace{-1mm}
\section{System Model}
\subsection {System Architecture and Eavesdropper Model}
We consider a mmWave wireless system consisting of two legitimate nodes, Alice and Bob, aiming to generate a shared secret key from their reciprocal wireless channel, while an eavesdropper, Eve, attempts to infer the key from overheard signals. Alice is equipped with an analog uniform linear array (ULA) of $N$ antennas with inter-element spacing $d$, while both Bob and Eve are equipped with a single antenna. \name's framework can be easily extended to support a multi-antenna Bob.\footnote{A multi-antenna eavesdropper may employ more sophisticated key reconstruction strategies depending on her hardware. Extending \name\ to account for such advanced adversarial models is left for future work.}

We assume that the environment, the nodes, and thus the wireless channel remain static during the key generation process; in other words, the channel coherence time exceeds the duration required to complete key extraction. Consequently, Alice must \emph{inject} randomness into the channel via her beamforming capabilities—otherwise, the resulting bit sequences would become repetitive and fail to meet randomness requirements. We focus on the downlink scenario, where Alice transmits and Bob receives, though the same principles apply to the uplink.

Prior to key generation, Alice and Bob perform an initial beam training phase in which Alice transmits a sequence of directional beams and Bob records the corresponding received powers. From this process, Alice acquires key channel information: the angular location of Bob (i.e., the LOS direction), the highest observed SNR at Bob, and the Rician K-factor. We assume that all three nodes operate in the same wireless environment and therefore, experience the same Rician K-factor.

We assume that Eve knows the exact locations of Alice and Bob and thus has full knowledge of the LOS path between them. However, Eve's actual position is unknown to Alice and Bob and may be arbitrary. As we show later, her most effective strategy is to position herself along the LOS and sample the transmitted signal in that direction to maximize her chances of inferring the key. To model a worst-case scenario, we further assume that Eve has extremely high receiver sensitivity, enabling her to detect arbitrarily weak signals—effectively operating under noiseless conditions.

Finally, we assume that the key generation protocol---that is, the set of public instructions used by Alice and Bob to derive the key---is fully known to Eve. However, Eve does not know the exact sequence of weight vectors selected by Alice during the key generation process, which prevents her from reconstructing the key even with full protocol knowledge.

\subsection{Channel Model}

We model the mmWave channel between Alice and Bob as a Rician fading channel comprising a deterministic LOS component and stochastic NLOS components. The normalized LOS component, denoted $\mathbf{h}_{\text{AB}}^{\text{LOS}}$, is a $1$-by-$N$ vector whose elements represent the phase shifts between each antenna at Alice and the single antenna at Bob.\footnote{Here, normalized means that each element has unit magnitude and that common complex scaling factors—such as path loss and single-element radiation pattern—have been factored out.} Assuming Bob is located at angle $\theta^{*}$ and applying the far-field approximation, $\mathbf{h}_{\text{AB}}^{\text{LOS}}$ reduces to the transmit steering vector at $\theta^{*}$, denoted $\mathbf{a}(\theta^{*})$: 
\begin{equation}
    \label{eq:h_los}
    \mathbf{h}_{\text{AB}}^{\text{LOS}} = \mathbf{a}(\theta^{*}) = 
    \left[
    1, 
    e^{-j k d \sin\theta^{*}}, 
    \dots, 
    e^{-j k (N-1) d \sin\theta^{*}}
    \right],
\end{equation}
where $j$ is the imaginary unit and $k$ is the wavenumber.

The normalized NLOS component, $\mathbf{h}_{\text{AB}}^{\text{NLOS}}$, is modeled as a $1$-by-$ N$ vector with i.i.d. complex Gaussian entries, each with unit variance. The complete channel from Alice to Bob is thus given by:
\begin{equation}
    \mathbf{h}_{\text{AB}} = 
    \sqrt{\frac{K}{K+1}} \mathbf{h}_{\text{AB}}^{\text{LOS}} + 
    \sqrt{\frac{1}{K+1}} \mathbf{h}_{\text{AB}}^{\text{NLOS}},
\end{equation}
where $K$ is the Rician K-factor, representing the power ratio between the LOS and NLOS components. The channel from Alice to Eve, $\mathbf{h}_{\text{AE}}$, is modeled in the same way, using the same K-factor, and includes both deterministic LOS and stochastic NLOS components. Recall that the K-factor and $\mathbf{h}_{\text{AB}}$ are estimated during the beam training phase, before key generation.

\subsection{Key Extraction Framework}
\label{sec:KeyGen}

As mentioned earlier, since the environmental channel is static, Alice must intentionally introduce randomness into the signal observed by Bob. To this end, in each time slot $t$, she applies a time-varying $N$-by-$1$ weight vector $\mathbf{W}(t)$ to her antenna array. The received signal at Bob, denoted $y_{\text{B}}(t)$, is then given by:
\begin{equation}
    \label{eq:y}
    y_{\text{B}}(t)=\mathbf{h}_{\text{AB}}\mathbf{W}(t) x(t) + z_{\text{B}}(t),
\end{equation}
where $x(t)$ is the transmitted signal (e.g., a known preamble) and $z_{\text{B}}(t)$ denotes AWGN at Bob. A similar expression applies to Eve's received signal.

The sequence of received signals at Bob over time must be quantized and mapped into a binary sequence using a predefined joint quantization and mapping rule. While multiple approaches exist, we adopt a differential phase-based scheme. Specifically, we extract the phase of the received signal $y_{\text{B}}(t)$ over $T$ time slots and compute the differential phase between consecutive slots. This approach mitigates discrepancies caused by unsynchronized local oscillators at Alice and Bob, which affect absolute phase readings. Moreover, phase is preferred over amplitude since amplitude fluctuations are generally more predictable—especially in LOS-dominant channels—and prior work has shown that amplitude-based schemes can be vulnerable~\cite{jana2009effectiveness}. Once the differential phases are obtained, they are quantized into $b$ bits per time slot, yielding a total of $(T\!-\!1)b$ random bits to be used as the secret key. The total number of bits, $(T\!-\!1)b$, should exceed the target key length, typically 128 or 256 bits.


The same key extraction procedure is independently performed by Eve using her own received signal. Since the key generation protocol is public, secrecy relies on ensuring that Eve’s observations remain uncorrelated with Bob’s. This is accomplished by intelligently shaping Alice’s beam pattern using a dedicated algorithm. To quantify the \emph{secrecy} provided by a given beamforming strategy, we use the key disagreement rate (KDR), defined as the number of mismatched bits between two keys divided by the key length. The KDR between Bob and Eve is denoted $\delta_{\text{BE}}$, and ideally, $\delta_{\text{BE}}$ should approach 0.5, indicating statistical independence between their keys.

On the other hand, secrecy is meaningless if Alice and Bob cannot extract matching keys. While channel reciprocity implies that they should ideally generate identical keys, noise introduces discrepancies that must be corrected during the information reconciliation phase~\cite{huth2016information}. High disagreement rates reduce the number of usable random bits and increase latency. Thus, achieving sufficiently high SNR at Bob is essential for minimizing mismatches—this is again facilitated by intelligent beam shaping, which governs how the beam pattern evolves over time. To evaluate \emph{robustness}, we use the key disagreement rate between Alice and Bob, denoted $\delta_{\text{AB}}$, which should ideally be close to zero.

\vspace{-1mm}
\section{Design}
\subsection{Naive Solutions}
\label{sec:naive}

Before introducing our beam-shaping algorithm for PLKG, we first demonstrate why conventional schemes fall short and motivate the need for a more deliberate design. Specifically, we examine two well-known baselines: random beamforming and null forming. In the first scenario, Alice selects beamforming weights randomly at each time slot, without considering the underlying channel, while Bob and Eve independently attempt to generate secret keys based on their received (noiseless) signals. We evaluate secrecy by measuring the KDR between Bob and Eve ($\delta_{\text{BE}}$) as a function of Eve's location. Additionally, we vary the Rician K-factor to study the impact of channel sparsity, with higher K indicating stronger LOS dominance—a common feature of mmWave propagation.

\begin{figure}[t]
    \centering
    \includegraphics[width=0.9\linewidth]{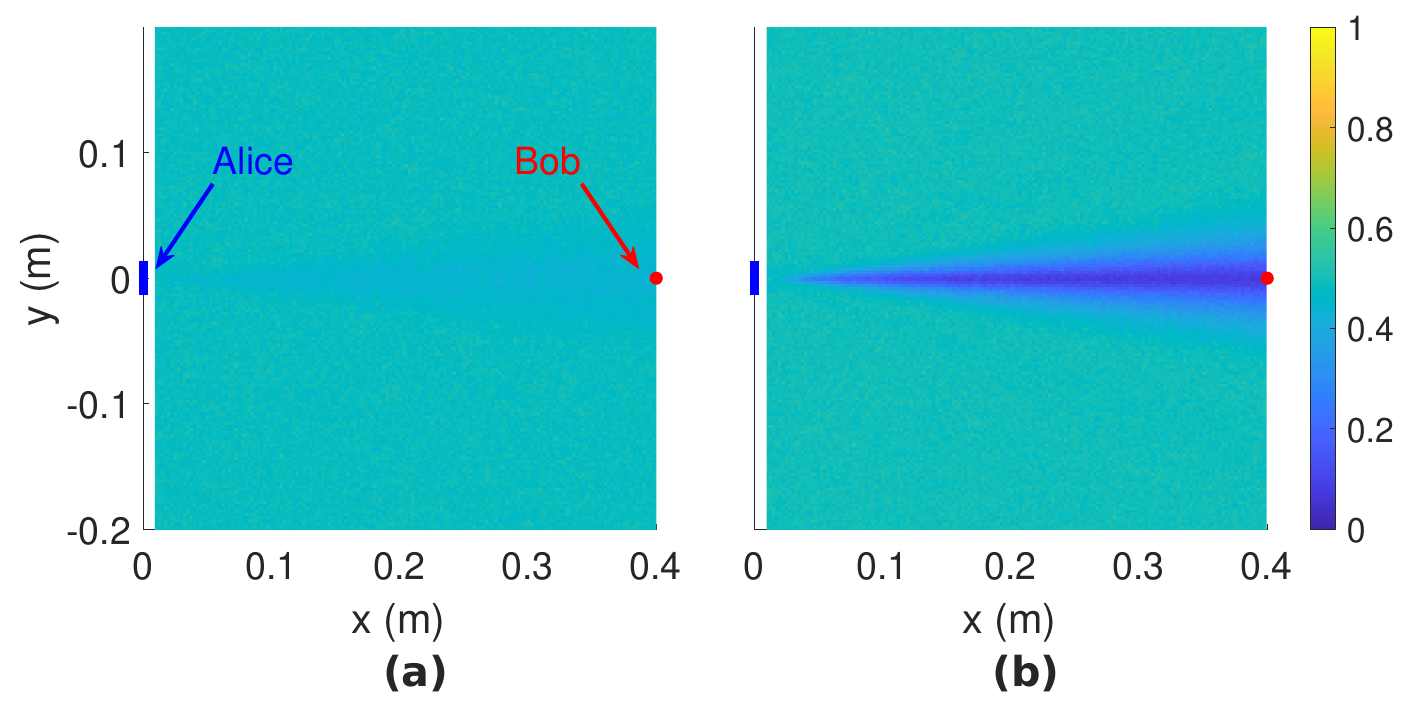}
    \vspace{-4mm}
    \caption{\textbf{The Limitation of Random Beamforming.} KDR between Bob and Eve ($\delta_{\text{BE}}$) under random beamforming, as a function of Eve's location for K-factor equal to (a) 0 dB and (b) 20 dB. In sparse channels, the eavesdropper can obtain a highly correlated key by observing the LOS.}
    \label{fig:random_bf_kdr}
\end{figure}

The results are presented in Fig.~\ref{fig:random_bf_kdr}. As expected, when the channel is rich in multipath (i.e., low K-factor), Eve’s observations remain uncorrelated with Bob’s due to the randomness introduced by reflected paths, as shown in Fig.~\ref{fig:random_bf_kdr}a. However, in sparser environments with higher K-factors, Eve can reconstruct a highly correlated key with Bob by positioning herself along the LOS path, as depicted in Fig.~\ref{fig:random_bf_kdr}b. In such cases, despite the use of random beamforming, the LOS component dominates the received signal at Bob, allowing Eve to infer the key simply by sampling the transmitted signal in the LOS direction.

\begin{figure}[t]
    \centering
    \includegraphics[width=0.9\linewidth]{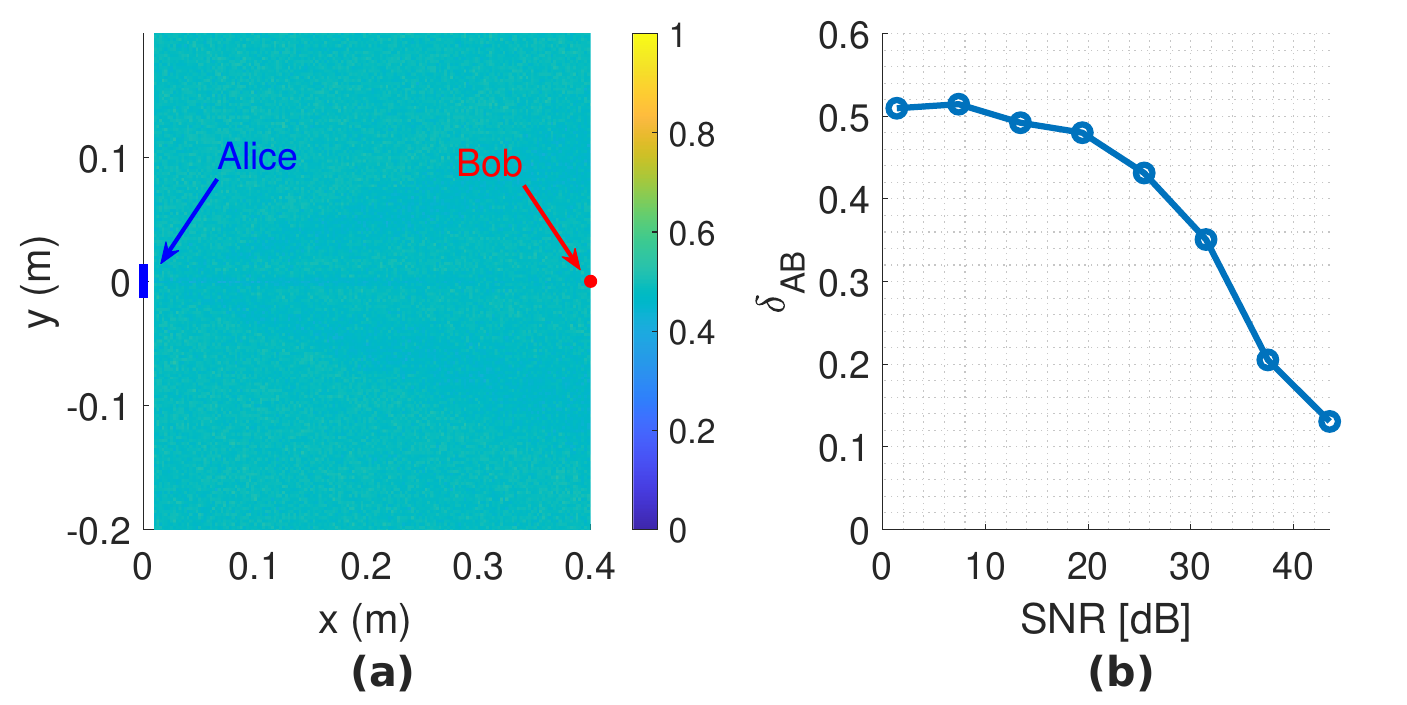}
    \vspace{-3mm}
    \caption{\textbf{The Limitation of Null Beamforming.} (a) KDR between Bob and Eve ($\delta_{\text{BE}}$) under null beamforming, as a function of Eve's location for K-factor equal to 20 dB. (b) KDR between Alice and Bob ($\delta_{\text{AB}}$) as a function of highest observed SNR during beam training. Even though nulling effectively provides secrecy, it compromises robustness by increasing $\delta_{\text{AB}}$.}
    \label{fig:nulling_effectiveness}
    \vspace{-5mm}
\end{figure}

Now that we understand Eve’s most effective strategy is to position herself along the LOS direction, a natural countermeasure is to null this direction to prevent her from sampling the key. To evaluate this approach, we repeated the previous simulation, this time selecting only weight vectors that produce a null toward the LOS. As shown in Fig.~\ref{fig:nulling_effectiveness}a, this strategy is indeed effective in degrading Eve’s channel—her KDR remains close to 0.5 across all positions, which is desirable. However, this method presents two key challenges:

First, if Alice had perfect continuous control over amplitude and phase, she could compute a weight vector that nulls the LOS direction by projecting onto the null space of the corresponding steering vector. In practice, however, limited resolution in both phase and amplitude control makes such nulling imperfect or infeasible, often resulting in residual leakage toward the LOS. Second, depending on the specific channel conditions, aggressively nulling the LOS can significantly degrade Bob’s received SNR. Since the LOS component dominates the received power in sparse mmWave channels, suppressing it weakens the legitimate link and increases $\delta_{\text{AB}}$, thereby compromising robustness.

This issue is illustrated in Fig.~\ref{fig:nulling_effectiveness}b, which plots the KDR between Alice and Bob ($\delta_{\text{AB}}$) as a function of the highest observed SNR at Bob during beam training. While this SNR serves as an indicator of the overall noise level, the actual SNR during key generation varies based on the beam-shaping strategy used. Under null beamforming, Bob suffers a substantial drop in effective SNR due to the intentional suppression of the LOS component. The figure shows that achieving acceptable $\delta_{\text{AB}}$ values under this strategy requires operating in a high-SNR regime—an assumption that may be unrealistic depending on factors such as transmit power and node separation.

\subsection{{\name}}
\label{sec:design}

Our main goal with \name\ is to strike a balance between the two key performance metrics in PLKG: robustness, measured by the KDR between Alice and Bob ($\delta_{\text{AB}}$), and secrecy, measured by the KDR between Bob and Eve ($\delta_{\text{BE}}$). These objectives are often at odds. At one extreme, aggressively nulling the LOS direction improves secrecy by limiting Eve’s ability to observe the transmitted signal, but this can significantly degrade Bob’s received power and reduce robustness. At the other extreme, random beamforming preserves signal strength at Bob and thus ensures robustness, but it exposes the system to LOS eavesdropping—especially in sparse mmWave environments.

There are various ways to balance robustness and secrecy depending on the application’s requirements. For instance, one may maximize robustness while ensuring that $\delta_{\text{BE}}$ remains above a minimum threshold. As a general-purpose strategy, however, we adopt the objective of maximizing the \emph{secrecy gap}, defined as
\begin{equation}
  \label{eq:sg_def}
  \text{SG} =
    \min(\delta_{\text{BE}},\,1-\delta_{\text{BE}})
    - \delta_{\text{AB}},
\end{equation}
which quantifies the difference between Eve’s and Bob’s key disagreement rates and offers a unified measure that captures both secrecy and robustness.

Intuitively, $\delta_{\text{BE}}$ depends on Eve’s position relative to the transmitter. Our design targets a worst-case eavesdropper who can position herself arbitrarily in the environment, with her location unknown to Alice and Bob. However, Eve’s most effective strategy is to align with the LOS direction and sample the transmitted signal along that path. Accordingly, \name\ is designed to ensure secrecy even under this strongest adversarial scenario.

To design a beam-shaping framework that maximizes the secrecy gap, we must first estimate the KDRs between Alice and Bob, and between Bob and Eve, for a given set of beamforming vectors. Recall that the Alice–Bob channel $\mathbf{h}_{\text{AB}}$ and the noise power at Bob are known from the initial beam training phase. Given a candidate set of weight vectors, \name\ estimates the instantaneous SNR at Bob using Eq.~\eqref{eq:y}, from which the expected value of $\delta_{\text{AB}}$ can be directly computed.

Estimating the average KDR between Bob and Eve is more challenging, as Eve’s precise location—and thus the exact Alice–Eve channel—is unknown. However, since Eve knows the positions of Alice and Bob, she can intercept their LOS path, and being in the same environment, her channel shares the same Rician K-factor. To model this, we generate samples of the NLOS component from a complex Gaussian distribution and combine them with the deterministic LOS component using the measured K-factor. This yields multiple realizations of the Alice–Eve channel, allowing us to estimate the expected $\delta_{\text{BE}}$ via Monte–Carlo simulation.

Now that we have established how to evaluate robustness and secrecy for a given set of beams, we define an optimization problem to identify the set of weight vectors that maximizes the secrecy gap. Unfortunately, this problem lacks a closed-form solution, and exhaustively searching the space of all feasible beams is computationally intractable due to the large design space. To address this, \name\ employs a genetic algorithm (GA) tailored to the constraints of Alice’s array—specifically, the per-antenna phase shifting resolution—to efficiently explore the beamforming space.

More specifically, the GA begins with a randomly generated population of weight vectors, each satisfying the total power constraint $\lVert\mathbf{W}\rVert^{2} = P_t$, where $P_t$ is the total transmit power. This population evolves over successive generations toward solutions that allocate less power in the LOS direction, as such beams improve secrecy. To guide this process, the GA minimizes the objective function $f(\mathbf{W})$, defined as the normalized LOS power:
\begin{equation}
  0 \;\le\;
  f(\mathbf{W}) 
  = \frac{\bigl|\mathbf{a}(\theta^{*})\mathbf{W}\bigr|^{2}}
         {N\,P_{t}}
  \;\le\; 1,
  \label{eq:los_obj}
\end{equation}
where $\mathbf{a}(\theta^{*})$ is the LOS steering vector (see Eq.~\eqref{eq:h_los}), $N$ is the number of antennas, and $\mathbf{W}$ is the beamforming weight vector applied at the transmitter.

Minimizing $f(\mathbf{W})$ promotes the selection of beamforming vectors that suppress the LOS component while adhering to the hardware constraints of Alice’s array. This evolutionary process continues until convergence, yielding a set $\mathbf{\Omega}$ of unique weight vectors with varying degrees of LOS suppression. To manage the secrecy–robustness tradeoff, \name\ applies lower and upper thresholds, $\alpha_l$ and $\alpha_u$, on the normalized LOS power $f(\mathbf{W})$, and selects a subset of vectors defined by:
\begin{equation}
  \mathbf{\Omega}' \;=\;
  \bigl\{
    \mathbf{W} \in \mathbf{\Omega}
    \;\bigl|\;
    \alpha_{l} \le f(\mathbf{W}) \le \alpha_{u}
  \bigr\}.
  \label{eq:omega_prime_recall}
\end{equation}

The values of $(\alpha_{l}, \alpha_{u})$ are selected by sweeping over candidate thresholds and evaluating the resulting $(\delta_{\text{AB}}, \delta_{\text{BE}})$ values. The pair that maximizes the secrecy gap is then chosen for key generation. Finally, during the key generation phase, Alice cycles through the beams in $\mathbf{\Omega}'$ across time slots, injecting controlled randomness into the otherwise static channel—enabling reliable key agreement with Bob while degrading Eve’s ability to infer the key, even under worst-case positioning.

\vspace{-1mm}
\section{Evaluation}
\vspace{-1mm}
We evaluate the performance of \name\ through simulations under various channel conditions representative of static mmWave environments. The setup assumes Alice is equipped with a ULA of $N = 9$ antennas spaced at half-wavelength and operating at 60~GHz. Each antenna features 2-bit phase control and 1-bit (on/off) amplitude control for beam steering. Both Bob and Eve are positioned at the array broadside.

The Rician K-factor for both the Alice–Bob and Alice–Eve channels is varied from 0~dB (rich scattering) to 30~dB (strong LOS dominance) to reflect diverse propagation environments. The highest observed SNR at Bob—measured using the best directional beam during beam training—is swept from 1.3~dB to 31.4~dB.\footnote{The highest observed SNR at Bob serves as an indicator of the environmental noise level. The actual SNR during key generation varies with the beamforming vector applied at each time slot and fluctuates with the selected beam pattern.}

Under each channel condition, we compare the performance of two baselines—random beamforming and LOS null beamforming (as defined in Sec.~\ref{sec:naive})—against our proposed framework, \name. The comparison is based on the KDR between Alice and Bob ($\delta_{\text{AB}}$) KDR between Bob and Eve ($\delta_{\text{BE}}$), as well as the secrecy gap (SG) defined in Eq.~\eqref{eq:sg_def}. When computing $\delta_{\text{BE}}$, we assume noiseless reception at both Bob and Eve to isolate the impact of beam shaping on secrecy from the effects of noise.

\subsection{Secrecy Gap}
\label{sec:eval_sg}
\begin{figure}[t]
    \centering
    \includegraphics[width=1\linewidth]{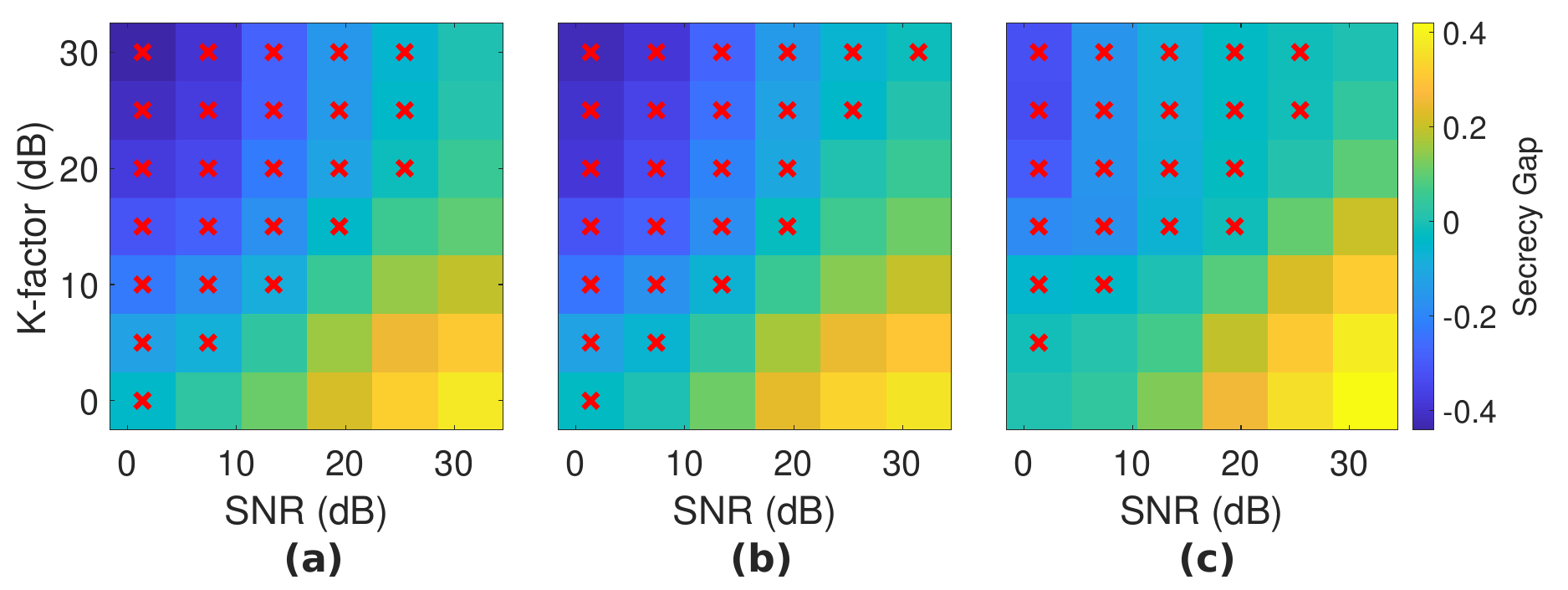}
    \vspace{-8mm}
    \caption{\textbf{Secrecy gap as a function of SNR and K-factor for (a) random beamforming, (b) null beamforming, and (c) \name.} Conditions where secrecy gap is negative are marked with a red cross, indicating channel conditions where PLKG should not be used.}
    \label{fig:SG}
    \vspace{-5mm}
\end{figure}

In this section, we compare the performance of \name~against the baselines in terms of the overall SG. The results are depicted in Fig.~\ref{fig:SG}. 

First, note that a secrecy gap below zero indicates that the eavesdropper can reconstruct the key more accurately than the legitimate node. In such cases, PLKG should not be used. Scenarios with a negative SG are marked with a red cross in Fig.~\ref{fig:SG}. This outcome occurs under our worst-case assumption of a LOS-aligned eavesdropper with extremely high receiver sensitivity, capable of detecting arbitrarily weak signals.

Another important observation from Fig.~\ref{fig:SG} is that the highest secrecy gap is achieved under rich multipath conditions (low K-factor) combined with high SNR. Multipath richness increases Eve’s KDR, enhancing secrecy, while high SNR improves KDR between Alice and Bob, reinforcing robustness. This trend is consistent across all evaluated schemes. In contrast, the most challenging scenarios for PLKG arise in sparse channels at low SNRs, where all approaches yield lower secrecy gaps. Nonetheless, on average, \name\ improves the secrecy gap by 39.4\% over random beamforming and 34.0\% over null beamforming.

\vspace{-1mm}
\subsection{Effect of SNR}
\label{sec:eval_snr}

\begin{figure}[t]
    \centering
    \includegraphics[width=1\linewidth]{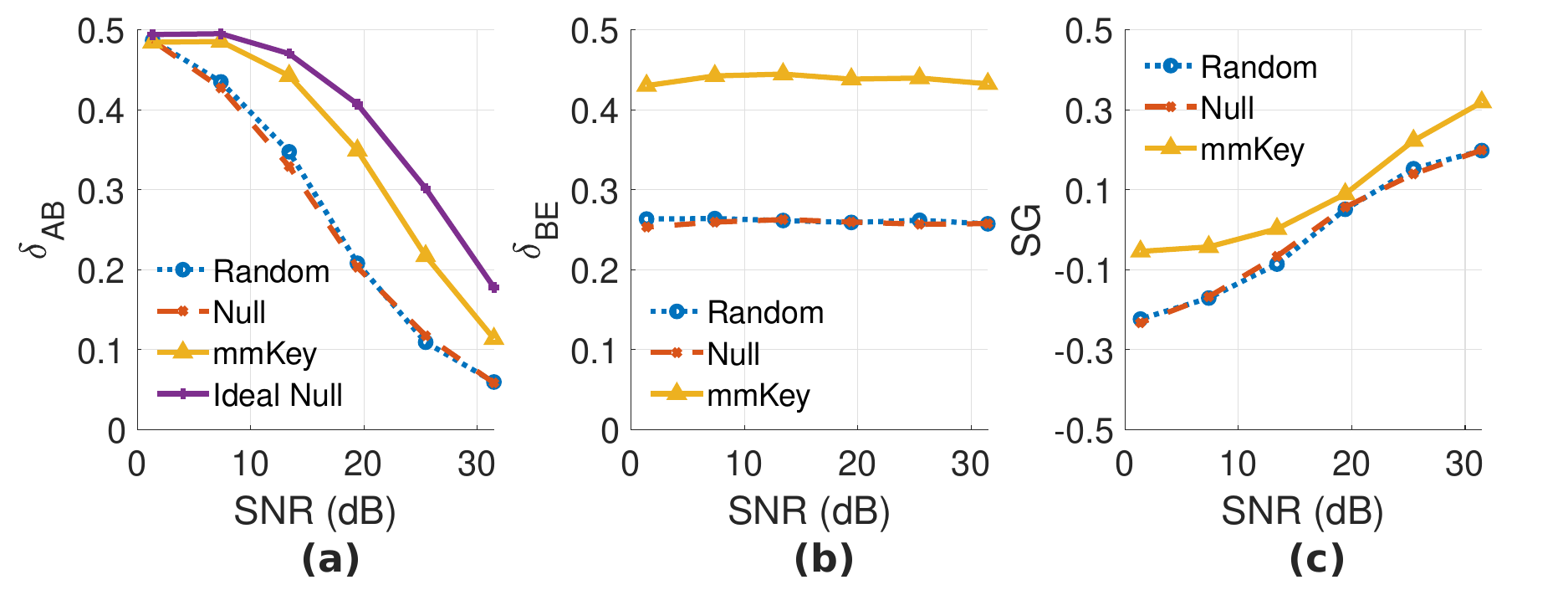}
    \vspace{-8mm}
    \caption{\textbf{Effect of SNR} on (a) KDR between Alice and Bob, (b) KDR between Bob and Eve, and (c) secrecy gap (SG) for different schemes when the K-factor is 10 dB.}
    \label{fig:SNR}
    \vspace{-5mm}
\end{figure}

In this subsection, we evaluate the impact of the highest observed SNR at Bob during the beam training phase on the performance of \name~compared to the baselines.  
The Rician K–factor is fixed at 10 dB, and the results are shown in Fig.~\ref{fig:SNR}.

As shown in Fig.~\ref{fig:SNR}a, the KDR between Alice and Bob, $\delta_{\text{AB}}$, decreases consistently with increasing SNR across all evaluated schemes—an expected trend, as higher SNR enhances channel reciprocity and improves key reliability. This figure also includes an ideal null beamformer with continuous phase and amplitude control, which results in a higher $\delta_{\text{AB}}$ than \name\ due to its aggressive suppression of the LOS component. In contrast, practical nulling—constrained by limited phase resolution—performs similarly to random beamforming. This is because computing continuous-domain weights and then quantizing them under a coarse resolution yields suboptimal nulling, making practical nulling largely ineffective and offering little benefit over random strategies.

Fig.~\ref{fig:SNR}b shows the secrecy performance of PLKG, measured by $\delta_{\text{BE}}$, across various beamforming schemes. As expected under the assumption of noiseless reception, $\delta_{\text{BE}}$ remains relatively constant across SNR levels. Nonetheless, \name\ consistently achieves higher $\delta_{\text{BE}}$ than the baselines, indicating reduced key leakage to Eve. This improvement stems from the fact that random beamforming does not suppress the LOS path, and while null beamforming attempts to do so, its impact is limited by coarse phase resolution. In contrast, \name\ employs a genetic algorithm to discover discrete weight vectors that effectively attenuate the LOS component, thus achieving superior secrecy.

Fig.~\ref{fig:SNR}c plots the secrecy gap as a function of SNR. We observe that the secrecy gap increases with SNR for all evaluated schemes, with \name\ consistently achieving the highest values. This trend occurs because, at higher SNRs, \name\ can allocate less power toward the LOS direction—reducing Eve’s ability to infer the key—while still maintaining a low KDR between Alice and Bob.

Finally, we note that \name\ exhibits slightly lower robustness compared to some baselines. This is expected, as its primary goal is not to optimize robustness in isolation but to maximize the overall secrecy gap. Consequently, \name\ strategically trades off some robustness (as seen in Fig.~\ref{fig:SNR}a) to achieve improved secrecy (Fig.~\ref{fig:SNR}b) and a higher overall secrecy gap (Fig.~\ref{fig:SNR}c).

\vspace{-1mm}
\subsection{Effect of K-factor}
\label{sec:eval_k}

\begin{figure}[t]
    \centering
    \includegraphics[width=0.78\linewidth]{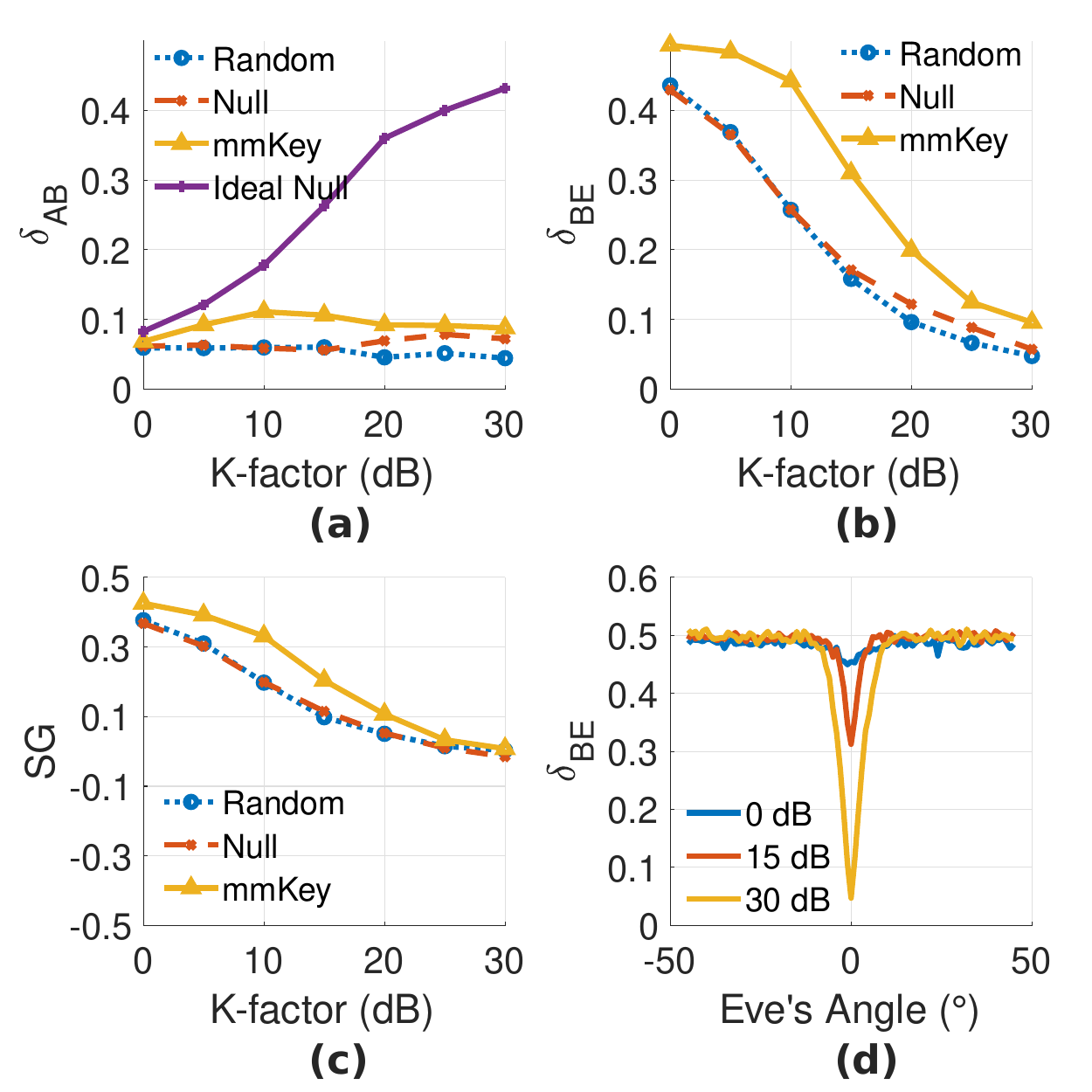}
    \vspace{-4mm}
    \caption{\textbf{Effect of K-factor} on (a) KDR between Alice and Bob, (b) KDR between Bob and Eve, and (c) secrecy gap (SG) for different schemes when the SNR is 31.4 dB. (d) Eve's KDR as a function of her relative angle to the LOS for different K-factors.}
    \label{fig:K-factor}
    \vspace{-5mm}
\end{figure}

In this subsection, we analyze the impact of the K-factor on system performance. The highest observed SNR during the beam training phase is fixed at 31.4~dB, representing the upper bound of SNR values typically seen in mmWave channels. The results are shown in Fig.~\ref{fig:K-factor}.

As shown in Fig.~\ref{fig:K-factor}a, the three main approaches maintain relatively stable values of $\delta_{\text{AB}}$ across varying K-factors. As in Sec.~\ref{sec:eval_snr}, we also evaluate the ideal nulling scheme, which shows a significant decline in robustness as the channel becomes sparser. This behavior is expected: in high-K environments, where the LOS path dominates, aggressively suppressing it—as in ideal nulling—substantially reduces Bob’s received signal strength, thereby increasing $\delta_{\text{AB}}$.

Fig.~\ref{fig:K-factor}b further shows that $\delta_{\text{BE}}$ degrades significantly with increasing K-factor. This trend aligns with Fig.~\ref{fig:random_bf_kdr}, where in sparse, LOS-dominant channels, an eavesdropper aligned with the LOS can more accurately observe the signal and reconstruct Bob’s key. Nevertheless, \name\ consistently achieves the highest $\delta_{\text{BE}}$ across all schemes, demonstrating superior secrecy performance.

As shown in Fig.~\ref{fig:K-factor}c, all schemes exhibit a decline in the secrecy gap as the K-factor increases. However, \name\ consistently outperforms the baselines by maintaining a larger secrecy gap, albeit with a slight reduction in robustness. Similar to the trend in Sec.~\ref{sec:eval_snr}, this reduction stems from \name’s design objective, which prioritizes maximizing the secrecy gap rather than minimizing $\delta_{\text{AB}}$. As a result, \name\ allows a controlled loss in robustness when it yields a higher overall secrecy gap—an outcome clearly demonstrated in Fig.~\ref{fig:K-factor}c.

Finally, we claimed that Eve’s best strategy is to position herself along the LOS and sample the signal in that direction to maximize her chances of reconstructing the key. To validate this claim under \name, we plot Eve’s KDR as a function of her angular offset from the LOS in Fig.~\ref{fig:K-factor}d across different K-factors. The results confirm that Eve achieves the lowest $\delta_{\text{BE}}$ when positioned at $0^\circ$, i.e., directly on the LOS.

\vspace{-1mm}
\section{Conclusion}
\vspace{-1mm}
In this work, we proposed \name, a PLKG framework tailored for static mmWave environments that leverages beam shaping to overcome the challenges of channel sparsity and LOS dominance. By using a genetic algorithm to iteratively search for weight vectors that suppress the LOS component while accounting for channel conditions such as sparsity and SNR, \name\ effectively balances the tradeoff between secrecy and robustness. Simulation results show that \name\ improves the secrecy gap by an average of 39.4\% over random beamforming and 34.0\% over null beamforming, demonstrating its effectiveness over conventional schemes.

\vspace{-1mm}
\bibliographystyle{IEEEtran}
\vspace{-1mm}
\bibliography{mmKey}

\end{document}